\documentclass[10pt,letterpaper]{article}
\usepackage[top=0.85in,left=2.75in,footskip=0.75in]{geometry}

\usepackage{amsmath,amssymb}

\usepackage{changepage}

\usepackage[utf8x]{inputenc}

\usepackage{textcomp,marvosym}

\usepackage{cite}

\usepackage{microtype}
\DisableLigatures[f]{encoding = *, family = * }

\usepackage[table]{xcolor}

\usepackage{array}

\newcolumntype{+}{!{\vrule width 2pt}}

\newlength\savedwidth

\raggedright
\usepackage[aboveskip=1pt,labelfont=bf,labelsep=period,justification=raggedright,singlelinecheck=off]{caption}

\makeatletter
\renewcommand{\@biblabel}[1]{\quad#1.}
\makeatother

\date{}

\usepackage{lastpage,fancyhdr,graphicx}
\usepackage{epstopdf}
\fancyheadoffset[L]{2.25in}
\fancyfootoffset[L]{2.25in}
\begin{document}
\vspace*{0.2in}

\begin{flushleft}
{\Large
\textbf\newline{A Critical-like Collective State Leads to Long-range Cell Communication in Dictyostelium discoideum Aggregation} 
}
\newline
\\
Giovanna De Palo\textsuperscript{1,2},
Darvin Yi\textsuperscript{3,4,\textcurrency},
Robert G. Endres\textsuperscript{1,2,*},
\\
\bigskip
\textbf{1} Department of Life Sciences, Imperial College, London, United Kingdom
\\
\textbf{2} Centre for Integrative Systems Biology and Bioinformatics, Imperial College, London, United Kingdom
\\
\textbf{3} Joseph Henry Laboratories of Physics, Princeton University, Princeton, New Jersey, USA
\\
\textbf{4} Lewis Siegler Institute for Integrative Genomics, Princeton University, Princeton, New Jersey, USA
\\
\bigskip


\textcurrency Current Address: Department of Radiology, School of Medicine, Stanford University, Stanford, California, USA 

*  r.endres@imperial.ac.uk

\end{flushleft}
\section*{Abstract}
The transition from single-cell to multicellular behavior is important in early development but rarely studied. The starvation-induced aggregation of the social amoeba Dictyostelium discoideum into a multicellular slug is known to result from single-cell chemotaxis towards emitted pulses of cyclic adenosine monophosphate (cAMP). However, how exactly do transient short-range chemical gradients lead to coherent collective movement at a macroscopic scale? Here, we use a multiscale model verified by quantitative microscopy to describe wide-ranging behaviors from chemotaxis and excitability of individual cells to aggregation of thousands of cells. To better understand the mechanism of long-range cell-cell communication and hence aggregation, we analyze cell-cell correlations, showing evidence for self-organization at the onset of aggregation (as opposed to following a leader cell). Surprisingly, cell collectives, despite their finite size, show features of criticality known from phase transitions in physical systems. Application of external cAMP perturbations in our simulations near the sensitive critical point allows steering cells into early aggregation and towards certain locations but not once an aggregation center has been chosen.

\section*{Author Summary}
Cells are often coupled to each other in cell collectives, such as aggregates during early development, tissues in the developed organism, and tumors in disease. 
How do cells communicate over macroscopic distances much larger than their typical cell-cell neighboring distance to decide how they should behave? 
Here, we develop a multiscale model of social amoeba, spanning behavior from individuals to thousands of cells.
We show that local cell-cell coupling via secreted chemicals may be tuned to a critical value, resulting in emergent long-range communication and heightened sensitivity. 
Hence, these aggregates are remarkably similar to bacterial biofilms and neuronal networks, all communicating in a pulse-like fashion. 
Similar organizing principles may also aid our understanding of the remarkable robustness in cancer development. 


\section*{Introduction}
Many living systems exhibit collective behavior, leading to beautiful patterns found in nature. Collective behavior is most obvious in animal groups with clear advantages in terms of mating, protection, foraging, and other decision-making processes \cite{Sumpter_2008,Vicsek_2012}. 
However, how cells form collectives without visual cues is less understood \cite{Mehes_2014}.
There are two main strategies to achieve synchrony (or long-range order) among individuals: firstly, a leader, i.e. a special cell or an external field, may influence the behavior of the others in a hierarchical fashion (top-down). 
An example is the developing fruit-fly embryo in maternally provided morphogen gradients \cite{Ashe_2006, Morrison_2012}.
Secondly, all individuals are equivalent and order emerges spontaneously by self-organization (bottom-up). 
Examples may include organoids \cite{VandenBrink_2014} and other cell clusters \cite{Kaliman_2014}. 
While order itself cannot be used to differentiate between the two mechanisms, the response to perturbations or simply the correlations among fluctuations can be examined \cite{Attanasi_2014}. In top-down ordering fluctuations are independent, while in bottom-up ordering fluctuations are correlated \cite{Cavagna_2010}.

To test these ideas of achieving order, we consider the well-known social amoeba {\it Dictyostelium discoideum}, which undergoes aggregation in response to starvation \cite{Levine_2013,Bretschneider_2002,Devreotes_1989}. 
During this developmental process, cells start to secrete pulses of cAMP, a molecule that also acts as a chemoattractant for the other cells in the vicinity. 
When a cell is `hit' by a high concentration of cAMP, it secretes a pulse of cAMP itself, relaying the signal and thus causing the formation of cAMP waves, inferred indirectly from optical density waves in dark field movies \cite{Rietdorf_1998, Weijer_1999}. These waves propagate through the whole population \cite{Alcantara_1974, Gross_1976, Tomchik_1981,Martiel_1987}. 
As their development proceeds, cells pulse at higher frequencies, reaching a maximal frequency of about one pulse every six minutes \cite{Gregor_2010}. 
Cell movement also accompanies the secretion process: 
before cells start to secrete cAMP, they normally move incoherently; when cAMP waves form, cells move towards the direction of the incoming wave by following the cells emitting the pulse in an orderly fashion (streaming phase). Interestingly, cells do not follow the passing wave in microfluidics (and hence solve the `back-of-the-wave' problem)\cite{Skoge_2014,Nakajima_2014}. 
While single-cell chemotaxis \cite{Skoge_2014,Nakajima_2014,Tweedy_2013,Meinhardt_1999,Parent_1999,Levchenko_2002} and large-scale pattern formation \cite{Kessler_1993,Levine_1994,Levine_1996, Maini_2012, Weijer_1999, Umulis_2013} have been extensively studied, a precise characterization of the transition from single cells to multicellularity is still 
missing.

Here, we develop a multiscale model to capture the mechanism of aggregation, focusing on the distinction between induced and self-organized order. Specifically, we are able to unify 
single-cell behavior and multicellularity at wide-ranging spatio-temporal scales. 
We achieve this by extending a single-cell model, which is able to describe {\it Dictyostelium} cell shape and behavior \cite{Tweedy_2013}, by adding intracellular cAMP dynamics, secretion, and extracellular dynamics for cell-cell communication. 
To simulate hundreds of cells, we extract a set of minimal rules for building a coarse-grained model. 
Hence, our approach is able to capture all stages of aggregation, ranging from single-cell chemotaxis to the multicellular collective.
For quantifying the transition from disorder (pre-aggregate) to order (aggregate), we employ spatial information and directional correlations. We found that the transition occurs during 
the streaming phase, which resembles a critical-like point known from phase transitions in physical systems using finite-size scaling arguments. Criticality and other predictions
are tested by corresponding analysis of time-lapse movies from fluorescence microscopy \cite{Gregor_2010, Sgro_2015, Noorbakhsh_2015}, pointing towards universal behavior in cell collectives.

\section*{Results}
\subsection*{A single-cell model fulfills criteria for aggregation}
\begin{figure}[!h]
\includegraphics[clip=true,trim=0cm 0cm 0cm 0cm,width=0.75\textwidth]{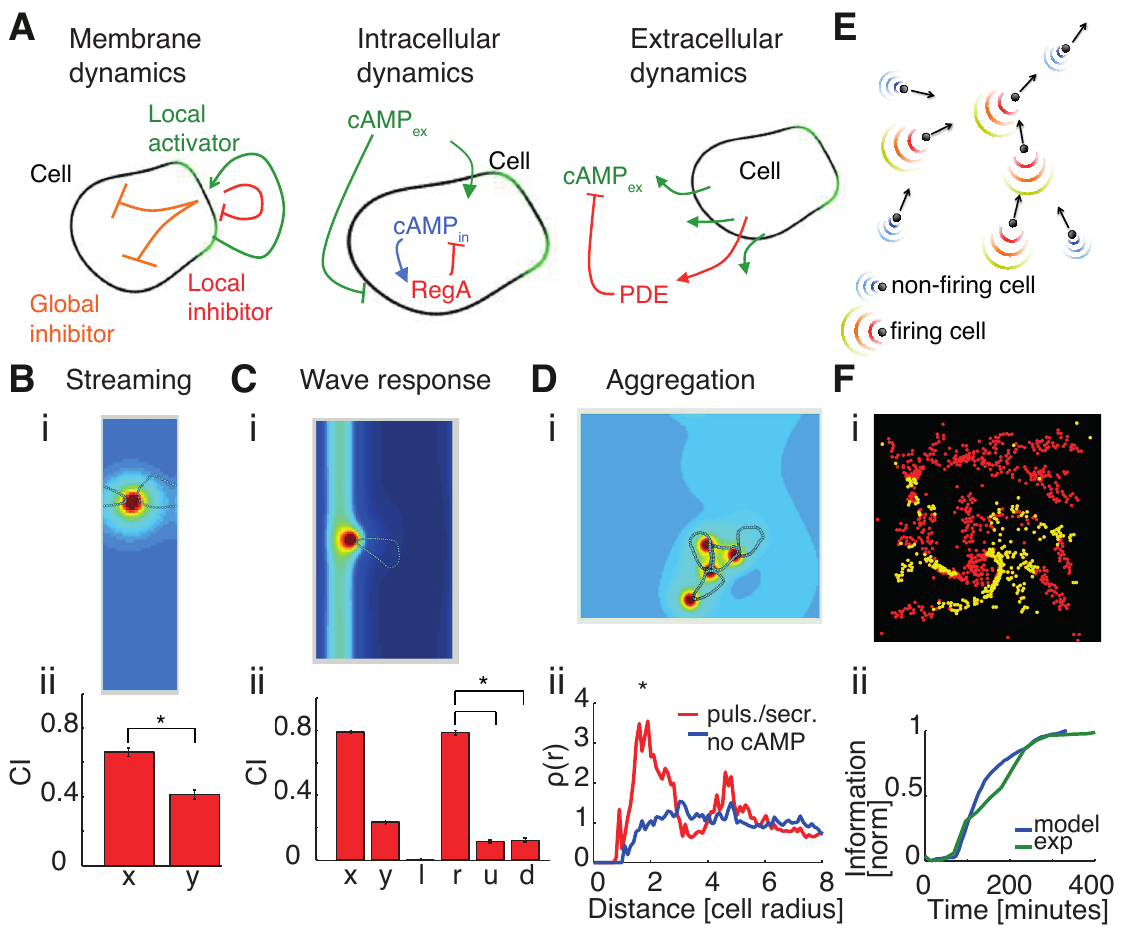}
\caption{{\bf Multiscale model: from single cell-shape changes and chemotaxis to collective behavior.} 
(A) Schematics for membrane dynamics (left), intracellular cAMP dynamics (center), and extracellular cAMP dynamics (right). 
(B) Single-cell ``streaming" simulation in a box with periodic boundary conditions and a constant concentration of cAMP (i).
Box dimensions are about 25x90 $\mu m$ (the initial cell radius is assumed $\sim 15 \mu m$). 
Due to the small dimension of the box the cell is just leaking, not pulsing, in order to avoid saturation of secreted cAMP. 
The simulation was repeated 12 times and the average chemotactic index (CI) calculated (ii). 
Errorbars represent standard errors. Differences in ${\rm CI}_x$ and ${\rm CI}_y$ are statistically significant (p$<$0.01), using a Kolmogorov-Smirnov test (KS test). 
(C) Cells solve `back-of-the-wave' problem.
(i) A Gaussian wave ($\sigma^2 \sim 60 \mu m$) moves from right to left with a speed of about 300 $\mu$m/min \cite{Tyson_1989}. 
At the peak of the wave, the cell emits a pulse of cAMP. 
After the firing, the cell enters a refractory period during which it can neither fire again nor repolarize. 
The cell generally moves to the right, and hence does not follow the passing wave.
(ii) CI in $x$ and $y$ directions, as well as for left, right, up and down directions in order to discriminate between the directions of the incoming (right direction) and outgoing (left direction) wave. 
Simulations are repeated 12 times; shown are averages and standard errors. 
Box is about 60x105 $\mu m$. 
CI in the right direction is significantly higher than CI in the other directions. 
(D) ``Aggregation simulation".
(i) Four cells are simulated moving in a constant concentration of cAMP. 
At the beginning, cells are randomly distributed. 
(ii) Density correlation at the end of simulations is plotted for control cells without secretion (blue) and all cells leaking cAMP and one cell also emitting pulses of cAMP (red). 
The red line has a significant (p$<$0.05, KS test) peak at a distance of about two cell radii, representing cell-cell contact. 
Also in this case simulations are repeated 12 times. 
Box dimensions are 75x75 $\mu m$.
See Materials and Methods for details on density correlation and {\it Supporting Information} for a full explanation of the detailed model.
(E) Schematic showing cells represented as point-like objects with velocity vectors. 
Firing cells emit pulses of cAMP, non-firing cells secrete cAMP at a low constant leakage rate. 
Spatial cAMP profiles are derived from detailed model simulations. 
At every time point cells are allowed two possible directions of movement in order to reproduce pseudopod formation at the cell front, with directions changing by $\pm 27.5^{\circ}$ with respect to the previous movement, corresponding to an angle between pseudopods of about $55^{\circ}$ \cite{VanHaastert_2010}. 
(F) Screenshot during streaming for $N$=1000 simulated cells (i). 
(ii) Spatial information versus time: simulations (blue) compared with experimental dataset 3 (green). 
Values were then normalized and shifted in time to facilitate comparison.} 
\label{fig1}
\end{figure}

To model the transition from single cells to multicellularity, we started with cell shape and behavior in single cells. 
Specifically, we considered a model capturing single-cell membrane dynamics similar to the Meinhardt model \cite{Tweedy_2013, Meinhardt_1999, Neilson_2011}. 
This model describes membrane protrusions and retractions, as well as resulting cell movement by means of three equations (see {\it Supporting Information}). 
The first and second variables are a local activator and a global inhibitor (both are also considered in the local-excitation global-inhibition -LEGI- model \cite{Parent_1999,Levchenko_2002}). 
The third is a local inhibitor, important to destabilize the current pseudopod and to increase the responsiveness of the cell (Fig.~\ref{fig1}A, left). To this we added dynamic equations representing the intracellular accumulation and secretion of cAMP from the cell rear \cite{McMains_2008, Kriebel_2008} based on the excitable FitzHugh-Nagumo model (Fig.\ref{fig1}A, middle) \cite{Sgro_2015,Noorbakhsh_2015}, as well as the extracellular dynamics for cell-cell communication (Fig.~\ref{fig1}A, right; see Materials and Methods for further information and {\it Supporting Information} for numerical implementation).

Using this detailed model, we investigated the resulting behavior of the cell-cell interactions in very small systems. 
First, we would like our model to capture streaming, i.e. the ability of a cell to precisely follow the cell in 
front of it. 
To reproduce that, we simulated a single cell in a rectangular box with periodic boundary 
conditions (see Fig.~\ref{fig1}B and supporting movie~S1). 
Given the rectangular shape of the box, a horizontally moving cell can sense its own secretion due to the periodic boundary 
condition. 
In contrast, a vertically moving cell is too far away from its rear and thus cannot sense its 
secretion. 
We estimated the ability of the cell to stream by measuring the chemotactic index (CI) in the 
$x$ direction, calculated as the amount of movement in the horizontal direction compared to the total length of the trajectory. 
In Fig.~\ref{fig1}B, we show that the CI in the $x$ direction is significantly higher than the 
CI in the $y$ direction.

We then considered the wave response as measured in microfluidic experiments, in which cells are exposed to traveling waves of cAMP \cite{Skoge_2014, Nakajima_2014}. 
When hit by a traveling wave, cells moved towards the direction of the incoming wave but 
did not follow the wave after it passed. 
In order to capture this behavior, our model cell undergoes a refractory 
period during which it cannot repolarize.
This is achieved naturally as the cell spontaneously emits a pulse of cAMP when encountering the wave (see Fig.~\ref{fig1}C, and supporting movie~S2). 
As a result, the CI is significantly higher in the right direction of the incoming wave. 
Finally, we considered a small number (four) of cells in a small box (with periodic 
boundary conditions) and tested whether they show signs of aggregation (see Fig.~\ref{fig1}D and supporting 
movie~S3). 
Specifically, we measured the density pair correlation (see Materials
and Methods), and compared the cases with and without secretion of cAMP. In the absence of secretion, cells 
were randomly distributed in space at the end of the simulations. With secretion, cells tended to be much 
closer to each other, with a clear peak in the density distribution at cell-contact distance (two cell radii).

\subsection*{A coarse-grained model reproduces collective behavior}

In order to reproduce aggregation as observed in experiments, e.g. \cite{Gregor_2010,Sgro_2015}, we need to 
simulate hundreds to thousands of cells. 
However, the detailed model introduced in the previous section is computationally too 
expensive, forcing us to introduce several simplifications. 
In our coarse-grained simulations, cells are point-like objects moving in continuous space. 
In particular, we took advantage of the spatial 
cAMP profiles from the detailed model by extracting the concentrations typically secreted by a single cell 
during leakage or a pulse, shaped by degradation and diffusion. 
As in the detailed model, the maximum 
cAMP concentration is always found in the direction opposite to the direction of motion (see Materials
and Methods). 
Using these analytical cAMP profiles, 
the cAMP concentration a cell senses is given by the sum of secretions by its neighboring cells. 
We then set concentration and gradient thresholds to determine whether a cell leaks or pulses cAMP, 
followed by a refractory period, and whether it moves randomly or follows the local cAMP gradient 
(see Fig.~\ref{fig1}E, Materials and Methods and {\it Supporting Information} for a detailed explanation of the model). 

Using this minimal set of rules, we simulated thousands of cells with a density similar to experimental ones
(around a monolayer -ML- \cite{Gregor_2010, Sgro_2015}). 
Cells were initially distributed uniformly in space and allowed to move randomly. 
As soon as the cell density 
increased spontaneously (and hence cAMP due to leakage by all cells), a cell may sense a concentration of cAMP 
large enough to pulse and this excitation will propagate throughout the whole population. 
Due to cell movement, 
streaming and aggregation into a small number of clusters can be observed (Fig.~\ref{fig1}F and 
supporting movie~S4). 
To quantify aggregation in a different way, we also calculated the
spatial information for capturing the order in an image based on the 2D Shannon information, relying on
Fourier coefficients while not requiring tracking of individual cells (see Material and Methods) \cite{Heinz_2011}. 
In all simulations, this spatial information rises sharply during the streaming phase as expected for cells in 
an ordered aggregate (see Fig.~\ref{fig1}G).
Interestingly, the spatial information was previously used to capture the second-order phase transition in the 
2D Ising model \cite{Heinz_2011}. Hence, we wondered whether aggregation may be viewed as a critical-like point?

\subsection*{Collective behavior: hierarchical or self-organized?}

\begin{figure}[h!]
\includegraphics[clip=true,trim=0cm 0cm 0cm 0cm,width=0.75\textwidth]{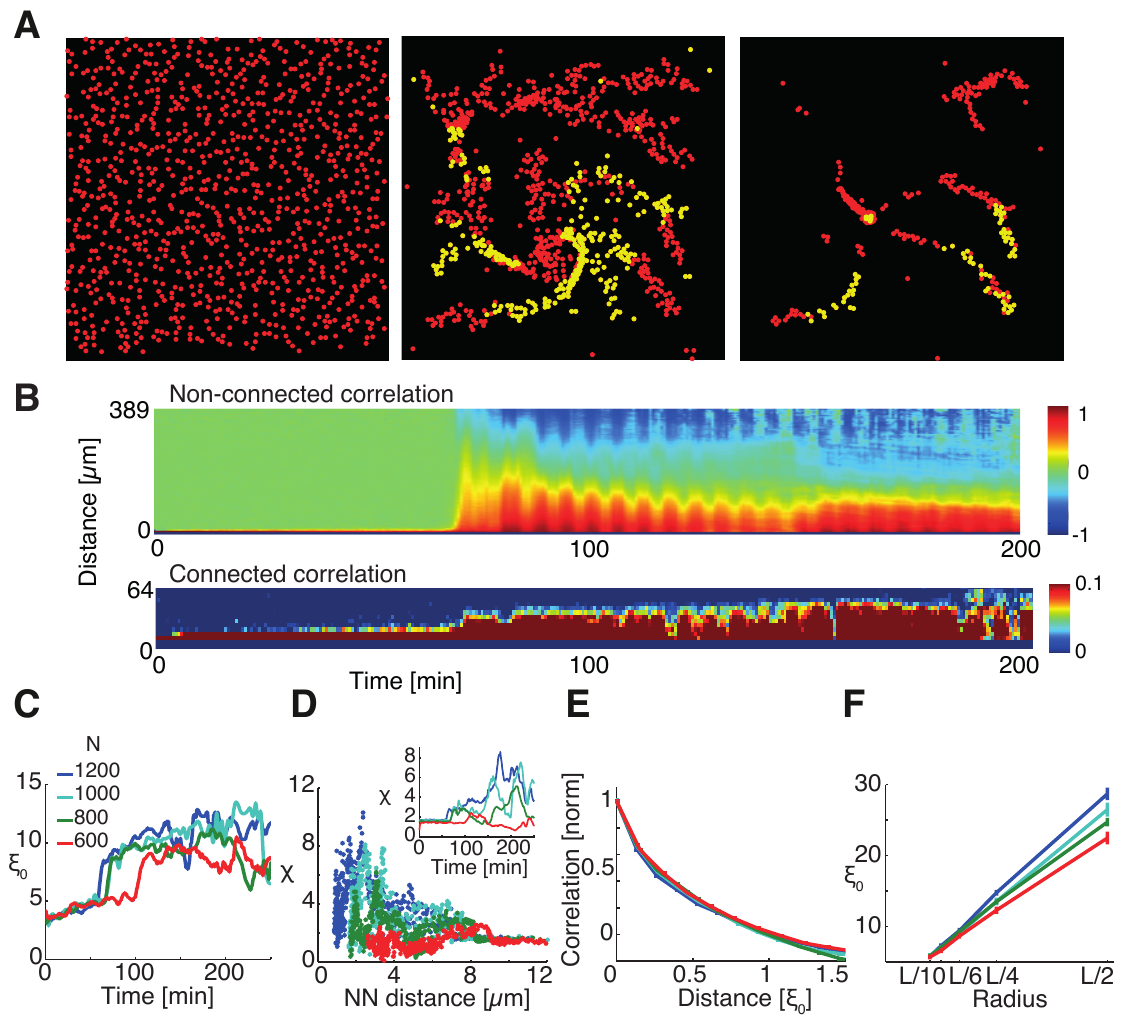}
\caption{{\bf Coarse-grained model leads to collective behavior.} 
(A) Screenshot of simulation for $N$=1000 cells at different time points: prestreaming (left), streaming (center) and after aggregation (right). 
Red (yellow) points represent non-firing (firing) cells. See {\it Supporting Information} for the full movie.
(B) Kymograph of non-connected $C_{nc}$ and connected $C_c$ directional correlations for simulation of $N$=1000 cells. 
Directional correlations profiles $C(r)$ were calculated for every time frame, and displayed depending on distance $r$. 
(C) Correlation length versus time for different numbers of cells.
Data were smoothed with a moving average filter spanning 10 consecutive frames. 
Inset: number of cells considered in simulations ($N$=1200, 1000, 800, 600).
(D) Susceptibility $\chi$ plotted with respect to nearest-neighbor (NN) distance for different number of cells and with respect to time (inset). 
The peak in susceptibility becomes the higher the larger the number of cells, and NN distances decrease accordingly. 
Profiles in the inset were smoothed with a moving average filter spanning 10 points.
(E) Comparison of correlation profiles for streaming phase (50-minute time window). 
Connected correlations were calculated for different numbers of cells and then normalized and plotted against their respective correlation lengths. 
The four profiles collapse onto a single curve, independently of the number of cells. 
(F) Average correlation length versus neighborhood radius for different number of simulated cells. $L$ corresponds to the size of the images (389 $\mu$m). 
$\xi_0$ represents the average correlation length during streaming phase (50 minutes window). 
Error bars represent standard error.
See {\it Supporting Information} for a full explanation of the model.
}
\label{fig2}
\end{figure}

Based on our model assumptions, all cells are treated the same. 
However, aggregation may still be driven by the
first random cell pulsing (hierarchical system) or spontaneously emerging as cells are coupled to
each other by cAMP sensing and secretion (self-organized system; Fig.~\ref{fig2}A). 
The order of the collective process can be measured 
studying the directional correlations of pairs of cells. Specifically, the {\it non-connected} ({\it nc}) correlations 
 \begin{align}
C_{nc}(r)=\frac{\sum_{i\neq j}^{N} \overrightarrow{u}_i \cdot \overrightarrow{u}_j \delta(r-r_{ij})}{\sum_{i \neq j}^{N} \delta(r-r_{ij})}
\label{eq:corr_nc}
 \end{align}
represent the average similarity of the direction of motion for every pair of cells depending on their distance, where $N$ is the total number of cells, $\overrightarrow{u}_i$ is the vector 
representing the direction of cell $i$, and $\delta(r-r_{ij})$ is equal to 1 if $r=r_{ij}$ and 0 otherwise.
$C_{nc}(r)$ also represents the order parameter in our system.
By calculating this quantity for every time frame, we can analyze its variation in time. 
During the preaggregation 
stage correlations are close to zero even at short distances, while they increase sharply during the streaming 
phase (Fig.~\ref{fig2}B, top).

Although order increases during the streaming phase, the origin and characteristics of this order are yet to be 
determined. 
To achieve this, we calculated the {\it connected} ({\it c}) directional correlations $C_c(r)$, measuring the 
similarity of the directional fluctuations with respect to the average velocity \cite{Attanasi_2014, Cavagna_2010}.
Thus, in this case direction $\overrightarrow{u}_i$ in Eq.~\ref{eq:corr_nc} is substituted by the velocity of the single 
cell when the average is subtracted, i.e. $\delta \overrightarrow{u}_i=\delta \overrightarrow{v}_i / 
\sqrt {\frac{1}{N}\sum_{j=1}^{N} \delta \overrightarrow{v}^2_j}$ with $\delta \overrightarrow{v}_i=\overrightarrow{v}_i-
\frac{1}{N}\sum_{j=1}^{N}\overrightarrow{v}_j$. 
For this kind of collective movement, such a subtraction is not straightforward. 
If we compute a {\it global average} velocity for every time frame, we systematically overestimate 
the non-connected correlations, because we still consider part of the ``bulk" velocity vectors due to the position of 
the cells in the image (see {\it Supporting Information} for a schematic explanation). 
To reduce this artefact, we consider {\it local 
averages}. 
For every cell, we consider the average velocity of all cells in its neighborhood up to a certain maximal 
distance $r_c$, and we compute the correlations between the cell in the center and all the cells belonging to its 
neighborhood, and repeat this procedure for every cell in our image.

When applied to the simulations, Fig.~\ref{fig2}B, bottom, shows significant connected correlations, especially during
streaming. Next, we considered the correlation length $\xi_0$ as the minimum distance at which the 
correlation crosses zero, i.e. $C(r=\xi_0)=0$ \cite{Attanasi_2014}. We find that $\xi_0$ is indeed 
much larger than the minimum nearest-neighbor distance, strongly suggesting self-organization (Fig.~\ref{fig2}C).

\subsection*{Streaming as a critical-like point}

Above, we demonstrated that aggregation in {\it Dictyostelium} is highly ordered and self-organized, with a correlation 
length much greater than the nearest-neighbor distance. Does the transition from disorder to order in this finite system show 
signs of criticality? 

In order to answer this question we considered that in critical systems the correlation length should scale with the size of 
the system as there is no intrinsic length scale \cite{Attanasi_2014}. 
To investigate this, we analyzed how the correlation 
length $\xi_0$ changes in time. In all simulations $\xi_0$ was small before aggregation and it increased markedly 
during the streaming phase (Fig.~\ref{fig2}C). The susceptibility can thus be computed as the maximum value reached by the 
integrated correlation
 \begin{align}
\chi= \frac{1}{N} {\sum_{i\neq j}^{N} \delta\overrightarrow{u}_i \cdot \delta\overrightarrow{u}_j \theta(\xi_0-r_{ij})},
 \end{align}
where $\theta(\xi_0-r_{ij})$ is equal to 1 for $r_{ij}<\xi_0$ and 0 otherwise \cite{Attanasi_2014}.
This proxy for the susceptibility peaks precisely during the streaming phase (Fig.~\ref{fig2}D, inset), and the higher the 
number of cells, the higher the susceptibility. Moreover, if we consider cell density as a control parameter (similar to 
temperature or coupling in a ferromagnetic Ising model), we can plot $\chi$ with respect to the rescaled nearest-neighbor 
distance (Fig.~\ref{fig2}D and Materials and Methods). 
The resulting peaks do not only follow the number of cells in terms of 
height, but also shift to smaller nearest-neighbor distances as the number of cells increases, further supporting the 
resemblance to a scale-free system near criticality \cite{Attanasi_2014}. 
Furthermore, normalizing the correlation and 
rescaling the distance by the correlation length, the correlations collapse for all our simulations when considering the average 
profile during the streaming phase (Fig.~\ref{fig2}E). 
Finally, we take advantage of our image partition with different 
radii $r_c$ to examine how the correlation length $\xi_0$ scales with system size. 
We notice that for all movies, higher cell numbers display longer correlation lengths for a given neighborhood radius, 
and that the correlation length increases with increasing radius.
As a result, the correlation length scales with system size (Fig.~\ref{fig2}F), indicating critical-like behavior in our simulated cells.

\subsection*{Analysis of time-lapse fluorescent microscopy}

\begin{figure}[h!]
\includegraphics[clip=true,trim=0cm 0cm 0cm 0cm,width=0.75\textwidth]{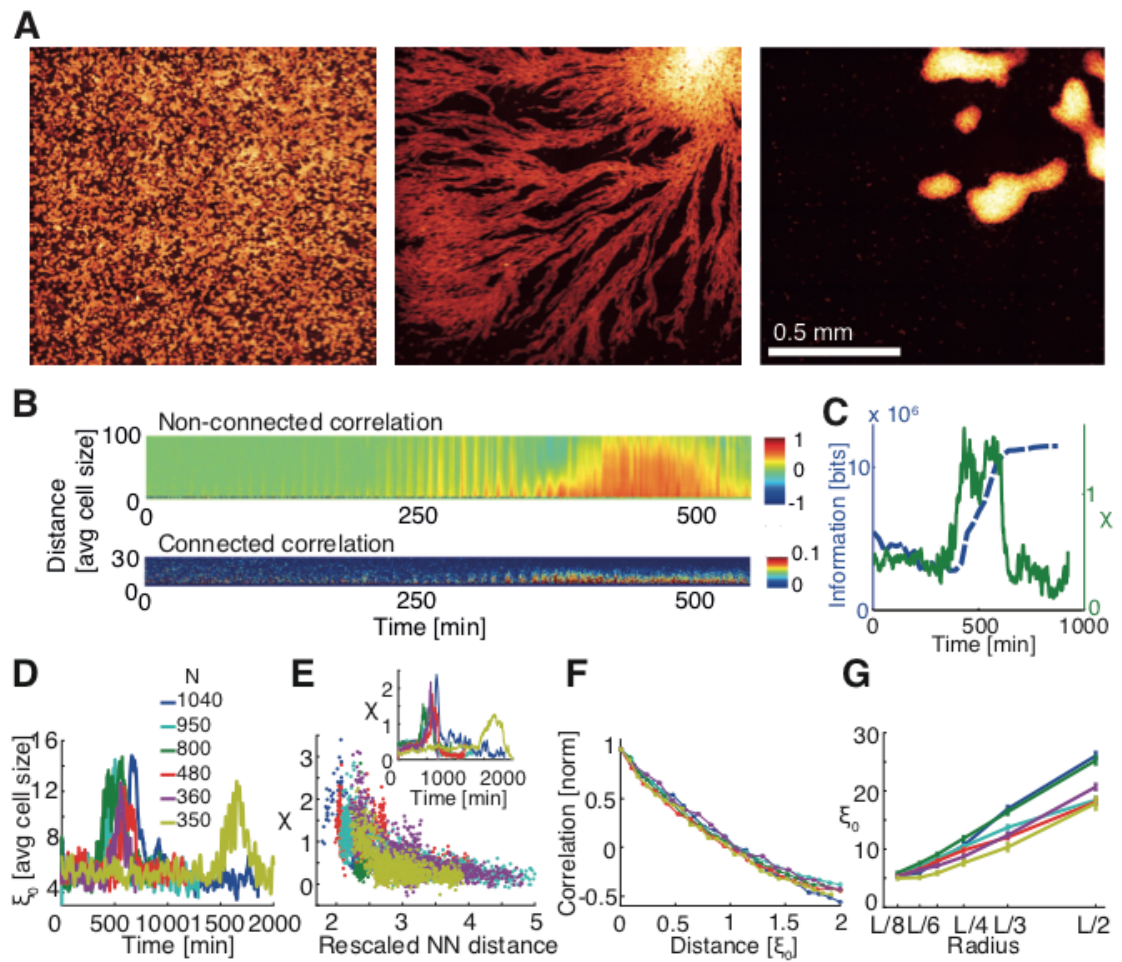}
\caption{{\bf Collective behavior in experimental data for validating model.} 
(A) CFP images of {\it Dictyostelium} aggregation of dataset 3. 
Images were taken after 4-5 hours of starvation when cells were still moving randomly before initiating aggregation (left), during streaming phase (450 minutes after first image, center) and after aggregation (800 min, right). 
(B) Kymograph of non-connected $C_{nc}$ and connected $C_c$ directional correlations for movie in dataset 3. Distance $r$ is expressed in units of average cell size (estimated after an ellipse was fitted to every cell contour, and corresponding to the average of the minor axis, $\sim 10.7 \mu m$). 
(C) Spatial information (blue) and susceptibility $\chi$ (green) of movie in dataset 3 as a function of time.
The increase in spatial information denoting a more ordered image corresponds to the peak in susceptibility. 
(D) Correlation length $\xi_0$ as a function of time for the six movies. 
Curves were smoothed with a moving average operation spanning 20 time points for better visualization.
Inset: comparison of cell number estimated from TRED images during streaming phase for different movies.  
(E) Susceptibility $\chi$ as a function of rescaled nearest-neighbor (NN) distance and as a function of time (inset). 
Note that height of peaks increases and that the corresponding rescaled NN distance decreases with number of cells, as for simulations. 
Rescaled NN distance was computed by normalizing NN distance by the average cell size.
In order to decrease noise, profiles in the inset were smoothed with a moving average spanning 20 time points. 
(F) Normalized $C_c$ as a function of correlation lengths $\xi_0$ for different movies. 
$C_c$ for every dataset was calculated as an average over 150 minutes of streaming phase. 
Error bars represent standard error. As in the simulated data curve collapse for different numbers of cells.
(G) Average correlation length versus neighborhood radius. 
$L$ corresponds to the size of images (2033 pixels, $\sim$ 1.3 mm). 
$\xi_0$ represents the average of 150 min during streaming phase. 
Error bars represent standard error.}
\label{fig3}
\end{figure}

To test the model, we analyzed five movies of {\it Dictyostelium} aggregation with different cell densities as described previously (see Materials and 
Methods and supporting movie~S5) \cite{Gregor_2010,Sgro_2015}. 
Briefly, during 15 hours of observation, individual cells become a single, multicellular organism, 
going through different stages including preaggregation, streaming and aggregation (see Fig.~\ref{fig3}A). Cell densities ranged 
from 1/3 ML to almost 1 ML, ensuring aggregation while restricting our system to 2D. 
A 10\% subpopulation of cells 
expression TRED fluorescent marker were tracked using a custom-written software (see Materials and Methods). 
Based on these cells,
we repeated the analysis from the simulated cells to the TRED cells from experiments, applying non-connected and connected correlations, 
correlation length, and susceptibility. 
As we used the same computational protocol for both simulations and data, a close
comparison was possible, allowing us to assess finite-size scaling and hence critical-like behavior. 

Based on this experimental protocol, we obtained five movies of different cell density, which changes slightly over the duration of 
observation due to open boundary conditions (see {\it Supporting Information}). 
Hence, cell numbers reported refer to the streaming phase (Fig.~\ref{fig3}D, inset). 
Based on our analysis, the correlation length $\xi_0$ increases 
during the streaming phase as does the susceptibility $\chi$ (Fig.~\ref{fig3}, B-E). 
Additionally, $\chi$ increases with cell number (and hence cell density), and the nearest-neighbor distance 
decreases, similar to the simulations. 
The correlation profiles, normalized and rescaled by the nearest-neighbor distance, 
largely superimpose for the different cell numbers, indicating that the slope of the resulting curves is not affected by the number of cells 
(see Fig.~\ref{fig3}F). 
Finally, we studied how the correlation length changes for different system sizes by considering different 
neighborhood radii as performed for the simulations (see Materials and Methods). 
We noticed that $\xi_0$ increases for a given radius 
with increasing cell numbers, and also for a fixed number of cells with increasing neighborhood radius (Fig.~\ref{fig3}G). 
These observations strongly suggest that there is no intrinsic correlation length, but that this length scales with system size. Taken
together, our results suggest that aggregation can be viewed as a critical-like point in this finite system.

\subsection*{Additional model predictions and cell `steering'}

\begin{figure}[h!]
\includegraphics[clip=true,trim=0cm 0cm 0cm 0cm,width=0.75\textwidth]{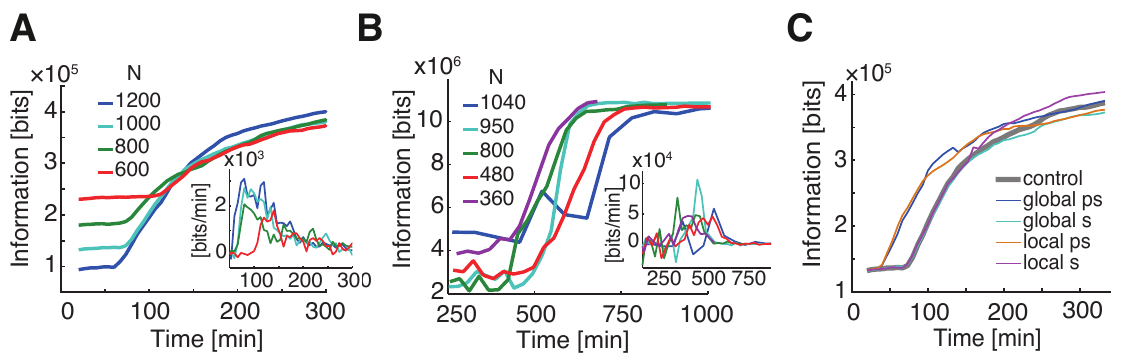}
\caption{{\bf Predictions from coarse-grained model.}
(A) Spatial information changes with $N$ in {\it in silico} data. Inset: Time derivative of spatial information profiles. The change in spatial information is larger for higher cell numbers.
(B) Spatial information as a function of time computed on the experimental data (as in panel (A), derivative is shown in the inset). As for the simulated data, the derivative tends to show higher peaks for experiments with higher cell densities.
(C) Effect of perturbations on the system during aggregation compared to control without perturbations. 
A speeding up of aggregation is seen if a localized or a global, spatially uniform pulse of cAMP is given to the system during prestreaming (ps). 
No effect on aggregation speed is noticed if the system is perturbed during streaming (s). See {\it Supporting Information} for the full movies (supporting movies~S6-S9).
}
\label{fig4}
\end{figure}

Here, we consider the dynamics of 
aggregation more closely. 
When increasing the density of cells in our simulations, we noticed that cells aggregate faster at higher cell densities as measured by the slope of the spatial information (Fig.~\ref{fig4}A). 
This increase in speed appears to reflect the increased ability to relay the signal as nearby cells can become excited and pulse themselves,
facilitating aggregation. 
To test this prediction, we attempted to quantify the speed of aggregation in our time-lapse 
movies as well. Although the experimental movies are much noisier and variable, we noticed a similar trend as in our simulation
(Fig.~\ref{fig4}B; only the dark blue curve violates the trend).

Critical-like behavior is the tipping point between order and robustness on one side and disorder and chaos on the other side.
This point may highten the sensitivity of the collective to detect changes in external cAMP concentration and to help cells make 
a decision on where to aggregate. 
While such experiments are beyond the scope of this work, we nevertheless attempted to investigate
this with our simulations. 
For this purpose, we applied both local and global cAMP perturbations to our system of cells; local perturbations represent a short local pulse of cAMP released on the 2D surface, while a global perturbation represents an overall
pulse of cAMP on the whole area of observation (see {\it Supporting Information} for details and supporting movies). In particular, we attempted to apply such 
perturbations before and during streaming,
with a noticeable different effect on aggregation. A local perturbation prior to streaming induces aggregation at the location of
the pulse, although the aggregation centre may shift later during the simulation. A global perturbation prior to streaming has a
similar affect, although the location of aggregation cannot be influenced. 
In contrast, applying such perturbations during streaming has largely no effect. 
While local perturbation may somewhat influence the location of aggregation, the overall dynamics stay the same. 
Indeed, comparing the time courses of the spatial information, we notice that the shape of the curve always stays the same, and only
the onset of aggregation can be shifted to earlier times with local or global cAMP perturbations (Fig.~\ref{fig4}C). 
In particular, early local
perturbations allow the possibility of steering cells to locations of aggregation.

\section*{Discussion}

{\it Dictyostelium} aggregation represents a fascinating example of synchronous collective cell behavior, spanning $\sim 1$mm
in length although cells are just $\sim10\mu$m in size. 
Here, we asked how cells produce such long-range communication \cite{Tweedy_2016}, 
when the transition from single cells to the collective occurs, and how this transition can be characterized quantitatively. 
To capture the main features of aggregation, we developed a multi-scale model. 
First, we focused on single cells using a 
detailed model combining sensing, cell-shape changes and movement with cAMP secretion/pulsing and hence cell-cell communication. 
Once this model resembled the behavior of a single cell or a small group of cells, it allowed us to extract a minimal set of rules 
that could lead to aggregation. 
In particular, we extracted the cAMP concentration profile of a pulse from the detailed simulations 
and the refractory period after pulsing. 
By allowing cells to leak cAMP and to randomly move below a certain cAMP threshold 
concentration, we were able to observe spontaneous random pulsing as soon as the local density increased, similar to what occurs 
in real cells. 
This minimal set was subsequently included in the coarse-grained agent-based model, which is able to reproduce the collective 
behavior of hundreds of cells in line with time-lapse microscopy \cite{Gregor_2010, Sgro_2015}. 

Our major findings point towards previously uncharacterized features in aggregation, both observable in simulations and
data. 
First, the transition to the collective is exactly pinpointed by a sharp rise in the spatial information of the cells during streaming. 
Second, to quantify the nature of the transition, we used fluctuations around the mean velocity, allowing us to distinguish between 
a hierarchically driven, top-down (external gradient from leader cells) and an emergent, self-organized, bottom-up (all 
cells are equal) process. 
Third, similar to second-order phase transitions in physical systems, the streaming phase shows signatures 
of criticality using finite-size scaling arguments. As a result there is no intrinsic length scale, allowing cells to communicate
with each other over large distances `for free', i.e. only based on local cell-cell coupling. 
The control parameter is cell density, affecting the cell-cell coupling via
cAMP secretion and sensing.

Our work provides further insights into the process of cell aggregation. 
By means of our multi-scale model, we were able to answer 
why cells emit cAMP in pulses. Albeit short-lived, a pulse creates a steeper spatial cAMP gradient than continuous secretion 
(assuming that the total amount of emitted cAMP is the same in both cases). Moreover, we noticed that so-called cAMP `waves' 
are likely not actual macroscopic traveling waves due to strong dissipation and diffusion. In contrast, cells are exposed to 
short-range cAMP pulses, which need to be relayed from one to the next cell before they dissipate. Although cAMP waves from 
microfluidic devices were used to study the cellular response to positive (incoming wave) and negative (passing wave) gradients, 
they may not represent natural stimuli \cite{Skoge_2014,Nakajima_2014}. Hence, cells may not have to solve the traditional 
`back-of-the-wave' problem, but instead have to decide which pulse to follow. However, this difficulty is eased as cells secrete 
cAMP from their rear \cite{Kriebel_2008}.

Our multi-scale model captures true emergence, generally not included in previous models of {\it Dictyostelium} aggregation. 
Models from the 1990s focused on the description of cell populations and the generation of spiral waves \cite{Kessler_1993,
Levine_1994,Levine_1996, Weijer_1999, Palsson_2000}. 
These were followed by the biologically more detailed LEGI \cite{Parent_1999,Levchenko_2002} and 
Meinhardt \cite{Meinhardt_1999} models to address the single-cell response to chemoattractant gradients. More recently, the 
FitzHugh-Nagumo model was adopted to explain the pulsing and synchronization of multiple cells (see {\it Supporting Information} for a comparison) 
\cite{Sgro_2015, Noorbakhsh_2015}.
Furthermore, hybrid models were proposed \cite{Bhowmik_2016}. 
However, none of these models started from a detailed spatio-temporal single-cell model and was able to describe the type of order and exact transition point for achieving collective behavior.

When dealing with complex biological phenomena, there are necessarily limitations in the deduced models and acquired data. 
To assess criticality via finite-size scaling, ideally cell density is varied by orders of magnitude. However, this is not 
feasible in this biological system. If cell density is much lower than about 1/3 ML, cells do not aggregate; if
higher, experiments would need to be conducted in 3D with major technical difficulties.  
Despite the approximations, our model allows the identification of the key ingredients for certain observed behavior. 
For instance, an earlier version of the model showed some level of aggregation but no finite-size scaling. 
By investigating this short coming, we noticed that streams were too narrow due to nearly negligible volume exclusion. 
However, quasi-one dimensional streams restrict cell movement and suppress criticality, reminiscent of the missing disorder-order phase transition in the 1D 
Ising model according to the Mermin-Wagner theorem \cite{Mermin_1966}.
(Note that the 2D Ising model is a borderline case, but it is still possible to formally define a phase transition according to Kosterlitz and Thouless \cite{Kosterlitz_1973}.) 
In our simulations, only when volume exclusion is increased and streams become broader, critical-like behavior emerges (see also discussion in \cite{Toner_1998}).    


In an attempt to unify wide-ranging biological phenomena, short-range interactions may play similar roles in cell collectives
({\it Dictyostelium}, neurons in the brain, biofilms, embryos, tumors) \cite{Moretti_2013, Chialvo_2010,Krotov_2014} and animal 
groups (such as bird flocks) \cite{Attanasi_2014, Cavagna_2010, Vicsek_1995, Nagy_2010, Bialek_2012}. 
Interestingly, neurons and bacteria pulse (spike) as well \cite{Prindle_2015}. 
Operating at criticality, i.e. the tipping point between order and disorder, may allow cells to be maximally responsive, to communicate robustly over 
long distances, to act as a single coherent unit, and to make decisions on, e.g., when and where to aggregate. 
In the future, it would be fascinating to conduct aggregation experiments in 3D environments, and to study the 
collective response to perturbations such as obstacles, changes in temperature, and exposure to toxins.

\section*{Materials and Methods}

\subsection*{Detailed model}
The intracellular cAMP dynamics are described by the FitzHugh-Nagumo model, a classical model to reproduce 
neuronal spiking and previously adopted to describe excitability in {\em Dictyostelium} \cite{Sgro_2015,
Noorbakhsh_2015}. Degradation of intracellular cAMP is achieved by phosphodiesterase RegA, which is negatively 
regulated by extracellular concentration of cAMP (by means of ERK2 \cite{McMains_2008}). Secretion of cAMP from 
the cell rear \cite{McMains_2008, Kriebel_2008} is strictly coupled to its intracellular concentration: if the 
extracellular cAMP concentration is below a threshold value cells exhibit a constant small leakage of cAMP, but 
a temporary high concentration of cAMP is released during pulses of intracellular cAMP once above the threshold. 
If the extracellular cAMP concentration is kept above this threshold the cell becomes a sustained oscillator. 
Extracellular cAMP is degraded by the phosphodiesterase PDE \cite{Bader_2007}. This model correctly captures 
the relay of the signal and the sustained pulsing observed in {\em Dictyostelium} (see {\it Supporting Information} for a detailed 
explanation).\\

\subsection*{Coarse-grained model}
To reproduce the dynamics of thousands of cells, we simplified further the representation given by the detailed model. 
We assumed that cells are point-like objects, which secrete cAMP maximally at their rear. 
Specifically, spatial propagation of cAMP was modeled as an exponential decay with a constant of 0.1 $\mu m^{-1}$ (within a factor of 2 of the value extracted from the detailed model simulations).
The spatio-temporal concentration profiles are rescaled according to the cosine of the angle with the opposite-to-motion direction; secretion becomes zero at 90$^{\circ}$ (lateral secretion) and is set to zero for all the frontal part of the cell.
(The above mentioned fine tuning of the exponential decay constant may be a result of this rescaling approximation, or may reflect the fact that the cell-cell coupling is a key parameter for critical-like behaviour.)
We set a concentration threshold $c_1$ to determine if a given cell will emit a pulse or just leak cAMP, and 
a gradient threshold $\nabla c_2$ determines if the cell will move randomly or follow the local cAMP gradient. 
As for the detailed model, every cell undergoes a refractory period of 6 minutes after firing, during which it keeps the same motion it had during pulsing.
To reproduce volume exclusion, cells cannot be closer to each other than 3$\mu m$ (this rule is overwritten later in simulations, when cells are densely packed and likely superimpose).
To drastically speed up simulations, the algorithm is written without explicit modeling of diffusion of cAMP in space, instead it computes how much cAMP every cell senses and what their spatial gradients are by considering positions of cells with respect to each other. 
This implementation is able to reproduce aggregation of thousands of cells.
More specifically, $N$=1000 cells were considered at experimental density of about one monolayer (6600 cells/$\mu m^2$). 
For the other simulations of $N$=600, 800 and 1200, the total area (of 389x389 $\mu m$) was fixed and density varied accordingly.
See {\it Supporting Information} for a detailed explanation.

\subsection*{Density pair correlation}
The pair-correlation function was computed as described in \cite{Gurry_2009}, given by
\begin{align}
g(r)=\frac{A}{N(N-1)} \frac{1}{2\pi r a}\sum_{i\neq j}^N \delta(r-r_{ij})
\end{align}
where $A$ is the total area of the image considered, $N$ is the number of cells, $r$ is the radius of a ring and $a$ is the discretization 
constant. In case of a random distribution $g(r)$ takes a value of 1 on average (similar to blue trace in Fig.~\ref{fig1}Dii), while 
in case of particle clustering $g(r)$ becomes greater for small distances (as for red trace in same panel). \\

\subsection*{Spatial information}
All images were binarized (by means of MATLAB thresholding algorithms {\em graythresh} and {\em im2bw} for the case of experimental images).
After that, 2D images were converted in 3D binary matrices where the third dimension has a 1 corresponding to the pixel intensity (thus in this case, since the starting images were binary, the 3D matrix has a 1 at level 0 if that pixel is black and at level 1 if it is white). 
This guaranteed that all images had the same histogram, provided that they initially were of the same size. For the case of 
uncorrelated pixels, all Fourier coefficients $P_i$ are considered independent and Gaussian distributed. Image entropies were then calculated 
as:
\begin{eqnarray}
H_{kS}=-2N\sum_i P_i\hspace{2pt} log_2 \hspace{2pt} P_i
\end{eqnarray}
where the probability density function $P$ is Gaussian distributed with zero mean and variance calculated from the sum of the 
pixel intensities. $H_{kS}$ is computed by dividing the function into bins of width $\sigma/100$ and summing $P_i\hspace{2pt} log_2 
\hspace{2pt} P_i$ from $-10\sigma$ to $10\sigma$. Fourier transformation was then applied to the image. The real and imaginary part 
of the Fourier coefficients were then considered to compute
\begin{eqnarray}
I_{kS}=\sum_i (-log_2 \hspace{2pt}P_i^R-log_2 \hspace{2pt}P_i^I)
\end{eqnarray}
where $P_i^R$ and $P_i^I$ refers to the real and imaginary part of coefficient $i$. The sum was calculated by considering bins 
of width $\sigma/100$ around the values assumed by the Fourier coefficients. $k$-space spatial information $kSI$ was finally 
calculated as $kSI=H_{kS}-I_{kS}$. For a more comprehensive explanation, see \cite{Heinz_2011}.
\\

\subsection*{Directional correlations and susceptibility}
To calculate the connected correlations, local averages of the velocities are subtracted from cell velocities. For every cell 
we considered the average movement of all cells in its neighborhood up to a certain maximal distance $r_c$, and compute the 
correlations between the cell in the center and all the cells belonging to its neighborhood. 
We repeated this procedure for every cell 
in our image. In this way we are able to decrease the ``bulk" velocity component in the fluctuations, while keeping a continuous 
partition of the image (which we would have lost in case of rigid partition of the image in smaller squares), and without preassigning 
the final position of the aggregation center. 
In order to understand better the influence of this partitioning on the calculation of 
the connected correlations, we repeated the same procedure for different radii. 
Specifically, if $L$ is the image dimension, we 
set $r_c$ equal to $L/2$, $L/4$, $L/6$, $L/8$, and $L/10$, with $L/6$ appearing to be the best choice in terms of the trade-off 
between avoiding overestimation of correlations and number of cells in the neighborhood for good statistics in the simulated data. 
For the analysis of experimental data, $L/2$, $L/3$, $L/4$, $L/5$, $L/6$, and $L/8$, were considered, and $L/4$ was chosen, reflecting again the trade-off between good statistics of noisy dataset and small overestimation of correlations.
To plot the susceptibility, we estimated the nearest-neighbor distance, computed for every frame as the average of the nearest-neighbor distances for all cells.

\subsection*{Experimental methods} Time-lapse movies were obtained similar to protocol in \cite{Gregor_2010, Sgro_2015}. 
Axenic {\em Dictyostelium} cells expressing Epac1camps were starved for 4-5 hours, and then plated on hydrophobic agar for imaging.
Sixteen fields of view from a microscope are combined (1.2 x 1.2$\,{\rm mm}^2$), resulting in the recording of thousands of 
cells in a wide field (inverted epifluorescence microscope (TE300, Nikon). To allow high-precision tracking of individual cells in a 
dense cell population, a different fluorescent marker (TRED) is expressed and mixed with unmarked cells so a subpopulation of cells could be tracked (10\% TRED cells).
See {\it Supporting Information} for further details.
\\
\subsection*{Segmentation and tracking}
Images of TRED channels were segmented by using the MATLAB function {\em imextendedmax}, which outputs a binary image given by 
the computation of the local maxima of the input image. The centroids positions were then computed from this mask by means of 
the {\em regionprops} function. The tracking of individual cells was done by considering the centroid positions for different times.
For every time $t$ the nearest neighbor centroid at time $t+1$ was found, and the trajectory was accepted if the distance between 
the two positions was smaller than the average cell size.
\\



\section*{Supporting Information}



\paragraph*{S1 Fig.}
\label{S1_Fig}
{\bf Coarse-grained model: cell sensing and behavior.}

\paragraph*{S2 Fig.}
\label{S2_Fig}
{\bf Schematic to explain choice of average in calculation of the connected correlations.}

\paragraph*{S3 Fig.}
\label{S3_Fig}
{\bf Oscillations in experiments and simulations.}

\paragraph*{S4 Fig.}
\label{S4_Fig}
{\bf Model responses to different stimuli.}

\paragraph*{S5 Fig.}
\label{S5_Fig}
{\bf Estimation of total numbers of cells.}

\paragraph*{S6 Fig.}
\label{S6_Fig}
{\bf Further support for critical-like behavior in the data.}

\paragraph*{S7 Fig.}
\label{S7_Fig}
{\bf Connected correlations for cell direction (top) and speed (bottom).}


 \paragraph*{S1 Video.}
{\bf Single-cell model: streaming.}
Single cell in a box to simulate streaming, see Fig. 1B of the main text.
 \label{SI_movie:det_streaming}

 \paragraph*{S2 Video.}
 {\bf Single-cell model: wave response.}
Single cell in a box to simulate the response to an external wave of cAMP, see Fig. 1C of the main text.
 \label{SI_movie:det_wave}

 \paragraph*{S3 Video.}
 {\bf Single-cell model: aggregation.}
 Four cells in a box to simulate aggregation, see Fig. 1D of the main text.
 \label{SI_movie:det_group}

 \paragraph*{S4 Video.}
 {\bf Coarse-grained simulations.}
 $N$=1000 cells, as shown if Figs. 1F and 2 of the main text.
 \label{SI_movie:N1000sims}

 \paragraph*{S5 Video.}
 {\bf Experimental movie.} 
 Dataset 3, shown in Fig. 3 of the main text.
 \label{SI_movie:exps}

 \paragraph*{S6 Video.}
 {\bf Global perturbation during prestreaming for coarse-grained simulations.} 
 $N$=1000 cells, see Fig. 1C of the main text.
 \label{SI_movie:globpert_prestr}

 \paragraph*{S7 Video.}
 {\bf Global perturbation during streaming for coarse-grained simulations.} 
 $N$=1000 cells, see Fig. 1C of the main text.
 \label{SI_movie:globpert_str}

 \paragraph*{S8 Video.}
 {\bf Local perturbation during prestreaming for coarse-grained simulations.} 
 $N$=1000 cells, see Fig. 1C of the main text.
 \label{SI_movie:locpert_prestr}

 \paragraph*{S9 Video.}
 {\bf Local perturbation during streaming for coarse-grained simulations.} 
 $N$=1000 cells, see Fig. 1C of the main text.
 \label{SI_movie:locpert_str}

\paragraph*{S1 Appendix.}
\label{S1_Appendix}
{\bf Supporting information.} Including detailed explanation of experimental procedure, single-cell and coarse-grained model and experimental data analysis.

\section*{Acknowledgments}
We are grateful to the Gregor lab at Princeton University for sharing their data with us, and additionally Thomas Gregor, 
Allyson Sgro, and Monika Skoge for helpful discussions. 
We also thank Luke Tweedy for help with the detailed model, Linus Schumacher for comments on the manuscript, and Mariam Elgabry and Suhail Islam for support with computational
issues. GDP and RGE were supported by ERC Starting Grant 280492-PPHPI (http://erc.europa.eu/starting-grants).


%
%
%

 \bibliographystyle{unsrt}

\bibliography{biblio}

%
%
%
%

\end{document}